\documentclass[letterpaper]{article}
\usepackage{aaai}
\usepackage{graphicx}
\usepackage{xcolor}
\usepackage{url}
\usepackage{times}
\usepackage{helvet}
\usepackage{courier}
\begin{document}
%
\title{MATILDA: Inclusive Data Science Pipelines Design through \\ Computational Creativity}
\author{
Genoveva Vargas-Solar, Santiago Negrete-Yankelevich, Javier A. Espinosa-Oviedo,\\ Khalid Belhajjame,  José-Luis Zechinelli-Martini\\
 Fundación Universidad de las Américas, Puebla\\
 Universidad Autónoma Metropolitana (Cuajimalpa) \\
%
 CNRS, Univ Lyon, INSA Lyon, UCBL, LIRIS, UMR5205,69622 Villeurbanne, France \\ 
 Université Paris Dauphine
}
\maketitle
\begin{abstract}

We argue for the need of a new generation of data science solutions that can democratize recent advances in data engineering and artificial intelligence for non-technical users from various disciplines, enabling them to unlock the full potential of these solutions. To do so, we adopt an approach whereby computational creativity and conversational computing are combined to guide non-specialists intuitively to explore and extract knowledge from data collections. The paper introduces MATILDA, a creativity-based data science design platform, showing how it can support the design process of data science pipelines guided by human and computational creativity.
\end{abstract}

\section{Introduction}
Exploiting large amounts of data collected in various domains through data science methods presents unprecedented economic and societal opportunities. Data science methods are primarily dominated by AI-savvy users with profound expertise in mathematical, statistical, numerical, and artificial intelligence models. However, these methods respond to research questions and challenges from other disciplines, with consumers with different expertise and little knowledge of data science methods.  Methods must go beyond the expertise of data scientists towards non-technical users with diverse expertise for them to be actionable. That is, engineering, humanities, social sciences, and any discipline requiring experiments on data to answer research questions. This new way of exploiting data calls for a new generation of data science solutions that harness and democratise recent advances in data engineering and AI for the benefit of users from non-technical disciplines, insulating them from technical complexity while enabling them to unlock the full power that these solutions can offer to meet the data analysis needs of their field.

The interaction and conversation of experts from various fields to address transdisciplinary problems call for creativity\footnote{Creativity is defined as the tendency to generate or recognise ideas, alternatives, or possibilities that may be useful in solving problems, communicating with others, and entertaining ourselves and others (\url{https://en.wikipedia.org/wiki/Creativity}).}  in designing data science solutions from different perspectives, namely, algorithmic, data, information technologies and trans-discipline non-data science users. Thus, there is a need for creativity-based solutions that can democratise data science that can guide non-specialists intuitively to explore data collections and extract knowledge from them.

The approach we investigate in this paper to address the above issues combines two strands, namely computational creativity, to allow users to explore uncharted territory when designing their data analysis, and conversational computing, to provide users with abstractions to steer and configure the data analysis without having to delve into low technical details.

Accordingly, the remainder of the paper is organised as
follows. The following section gives a general overview of approaches contributing to model creative-based processes and friendly design data science pipelines. It discusses how both areas can provide novel ways of designing data science-driven solutions. The section after that describes the challenges of modelling creative-driven data science design processes. It gives the
general lines of an approach that can enhance and envision a new way of addressing analytics problems using data and artificial intelligence models. The following section after that introduces a creativity-based data science design platform. It describes the general
architecture and functions and shows how it can support the design process of data science pipelines guided by human and computational creativity. The final section concludes the paper and discusses future work.

\section{Related work}\label{sec:relatedwork}
Addressing the creative design of data science pipelines to make them inclusive for non-experts requires drawing from methods and results from three areas: creative models, friendly data science and provenance. This section gives an overview of the relevant results that provide a scientific background to the project and help to contextualize our objectives.

 \it{\bf Creativity-driven systems}
Artificial intelligence (AI) offers opportunities to reify and transform how we think about human cognitive capabilities. In this context, computational creativity (CC)\footnote{Computational creativity is the study of building software that exhibits behaviour that would be deemed creative in humans. Such creative software can be used for autonomous creative tasks, such as inventing mathematical theories, writing poems, painting pictures, and composing music (\url{https://computationalcreativity.net}}, aims at studying human creativity and building systems that perform in such a way as to be considered creative. Concerning creativity, whether this cognitive capability is individual or collective. CC has moved from the classic individualistic and cognitive model of creativity \cite{boden2004creative} to a social and collective creativity model \cite{maher2010evaluating,maher2012computational,negrete2014apprentice,sawyer2011explaining,goel2010special,gu2021critical}. 

Collective-creativity models are concerned with understanding the roles and tasks that different agents (both human and non-human) play in a process that requires creativity and how creativity can be measured in this context \cite{kantosalo2015interaction,kantosalo2019experience}. Co-creativity is an excellent approach to establishing efficient, context-aware interactive systems and setting long-term programmes where solutions to problems requiring human and machine collaboration can be studied.

CC has been widely applied in art with systems that promote “artificial” creation. 
``The painting fool" is a system developed by Simon Colton that draws portraits taking into account emotional information obtained from the subjects being painted through a camera \cite{colton2011computational}.
Negrete-Yankelevich and Morales-Zaragoza \cite{negrete2014apprentice} propose a model to develop and assess creativity in computational agents embedded in mixed teams. The Apprentice Framework model establishes a series of roles (or levels of responsibility) agents can play within the group over time with the possibility of ascent through the ladder as the system is developed, acquiring thus more responsibility in the creative process. The model also helps identify aspects of the product being produced, at which point the agent is supposed to be creative. By keeping track of both responsibilities and aspects, it is possible to plan and assess the development of the system.
Negrete-Yankelevich and Morales-Zaragoza \cite{negrete2013motion}  propose a framework to create animatics. These animated storyboards constitute an essential artefact in producing animations by a team of expert animators that made an award-winning series of one-minute shorts for Mexican TV called Imaginantes (Televisa, ``Imaginantes* - YouTube."). The system's creativity is measured by how well the overall creativity of the team is affected by the system's performance.

\it{\bf Developing friendly data science solutions. }
Friendly data science systems must provide intuitive and interactive access to data processing operations in an agile and visual step-by-step manner \cite{bethaz2021ds4all}. They should help a user to derive conclusions about the data collection content and identify the potential questions that data can help answer. Through conversational loops and feedback, a friendly exploration and analysis system must calibrate the tasks according to the data's characteristics and the user's expertise and expectations. Through metadata collection and user profiling, an exploration and analysis conversation loop should propose actions, insight, and results' display (and visualisations) that assist the user in completing a given goal.

\it{\bf Discussion.}
We believe the collective CC can reproduce the collaborative transdisciplinary conditions in which data science solutions are developed.  It can drive the proposal of data science solutions combining data, algorithms, and computing resources to model complex systems and contribute to answering research questions to understand and predict them. 
While computational creativity and conversational techniques have proven effective, they have not been explored with the design of exploratory data analysis pipelines. Moreover, the two approaches are somewhat opposed because conversational techniques tend to rely on known territories (i.e. previously explored data manipulation and analysis actions). In contrast, computational creativity allows for exploring unknown territories (data manipulation and analysis), which may, in some cases, prove more effective. 

Our work addresses two challenges. On the one hand,  adapt and leverage both techniques to design an efficient and exploratory data analysis pipeline. On the other hand,  strike the right balance when creating data analysis pipelines between 'known' prior data exploration and analysis actions and 'unknown' creative actions.

\section{Creative process for designing data science driven solutions}\label{sec:approach}
Data science pipelines combining machine learning and deep learning are the new query types with specific needs regarding how data must be structured and managed. The “one all-fits-all” data structure and associated management functions approach are no longer adapted for data science queries. Indeed, every query has a specific objective (modelling, prediction), and its design entirely depends on the input dataset and an initial research question (RQ). The data science query is not based on an explicit knowledge of the data. It includes tasks devoted to mathematically understanding the data; then, the partial results of those tasks determine the design of other studies devoted to the computation of a model representing some hidden knowledge. Given statistical and machine learning methods and a target objective, data scientists rely on libraries that provide methods that they combine to define a data science pipeline. The results obtained by this pipeline are never definite, and they are always, to some degree, close to the target. 

To illustrate the design process of a DS pipeline, consider the following scenario. Consider a trendy decision-making group willing to adopt a data-driven approach for designing public policies to enhance citizens' lives in urban spaces and reduce energy and economic costs. Public policies are intended to modify built environments to improve them from financial and well-being perspectives. Decision-makers know that
from the urbanism perspective, small changes in the built environment can alter how people use the space. For instance, increasing pedestrian areas in a city downtown close to restaurant zones reduce CO2 footprint. Still, it impacts the influx of restaurant customers in the area and lowers real estate prices. Customers can suddenly start preferring restaurants close to parking slots. People living in the area can have problems accessing it and park their cars close to home. The research question is {\em to which extent public policies can impact the quality of life of different categories of citizens willing to evolve in a given urban area?}. Decision makers now call for data scientists' creativity to provide studies and mathematical evidence of the kind of urban changes to be considered in public policies. 

\it{\bf Designing a data science pipeline.}
Datasets and research questions drive DS pipelines' design. Through a  simplified creative scenario using the main phases of a DS pipeline:  (1) collect or search for datasets that can be used for answering a research question, and then (2) prepare them (explore, clean, engineer) to feed one or several Artificial Intelligence (AI) models. (3) Models are trained and tested with dataset fragments. These tasks are calibrated recurrently until specific performance scores are reached. (4) Results are constantly assessed and eventually considered good enough to be interpreted by experts and conclude on answering the initial research question to some extent.

Sketching a creative process, the elements to consider are: What data is needed to answer the research question and develop a strategy to collect them? How to transform the initial research question into a quantitative statement that can be addressed by mathematical or AI models? Which models' families can be pertinent for answering the question? How to design a series of tasks where data are processed? How to connect the results format with the research question statement? How to determine whether results converge? How to decide whether results are fair enough for considering an answer?

In the example, data scientists can decide to film civilians in the target urban spaces to collect their behavioural patterns on how they occupy and evolve along those spaces before and after implementing public spaces. Extract behavioural patterns that imply designing a DS pipeline for processing videos and detecting civilians, for example, using perceptrons \cite{cruz2022examination} and behaviour patterns within a series of scenes. The patterns can then be classified according to properties that detect changes before and after implementing some change. Other possibilities would be to run other data collection techniques like questionnaires to describe urban civilians' behaviour through quantitative variables that can be correlated for detecting changes produced after applying public policies. The possibilities are numerous, and they rely on data scientists' expertise, on the facilities or not for collecting certain types of data (e.g., video vs questionnaires) and their knowledge of specific AI models' families.





\it{\bf Discussion}
Data Science and Machine Learning Environments provide all the necessary AI models. They are supported by enactment stacks that deal with the storage, fragmentation, indexing and distribution of the data required and produced by the tasks composing a pipeline. What are the rules and strategies to combine different components that can transform input data into models and predictions that provide quantitative elements to answer initial research questions? 

Generative artificial intelligence \footnote{According to the Bing chat-GPT and validated by this paper's authors: {Generative AI refers to a category of artificial intelligence (AI) algorithms that generate new outputs based on the data they have been trained on (\url{www.weforum.org/agenda/2023/02/generative-ai-explain-algorithms-work/})}.} has started to be consolidated into solutions that give the illusion of creation through interactive and conversational approaches \footnote{Microsoft Deepspeed \url{https://github.com/microsoft/DeepSpeed/tree/master/blogs/deepspeed-chat}}. Systems like chat-GPT, in its various versions, mimic conversational and question-answering experiences intended to perform target tasks or produce ``new" content based on existing evidence. The principle of this system is synthesising the creation process as an exercise of wrapping together ``content"  with specific characteristics and considering some constraints to produce artefacts that look, to some extent, novel. 

In the case of DS pipelines, the first challenge is to model the creation process behind them. How does someone (a domain expert) states a research question such that a data-driven quantitative study can be run? How are data collected and selected to answer such questions? Which comes first, data or questions? How to conclude that given datasets representing observations of an object of study are representative enough to produce a model or predict the behaviour of that object? How is the human integrated into the loop and intervenes in the design milestones of a DS pipeline?

\it{\bf Challenges and Open Issues.}
A computational-creativity based methodology for designing DS pipelines should consider at least the following  scientific challenges and associated open problems:\\
    - Modelling hybrid (human and nonhuman) creativity-driven data science pipelines' design:  propose a  computational creativity model to represent end-to-end pipeline design. The creativity model can integrate design patterns like the ones presented in \cite{glines2021software} (design, mutant shopping, chorus line, simulation and approximating feedback, entertaining evaluations and no blank canvas). Depending on the tasks to be designed within a DS pipeline, different creativity patterns can best be adapted to address the task. \\
   %
 - Define the interaction among humans designing a DS pipeline and artificial system(s) that can take on tasks and propose results. Model the input/out required to feed and expect to/from the system and the type of feedback to be given by humans.\\
 - {\it Intervene} the process with an agent by selecting a relevant subprocess where creativity would contribute significantly to the overall solution and assess how it works. Then try other similar subprocesses and verify again. This bottom-up approach would establish a practice to turn the overall process into a friendly one in a stepwise fashion.\\
 %
   - Collecting provenance and data from DS pipelines design tasks: implement processes for data curation, annotation, identification, and quality control in research.\\
%
    - Proposing an ad-hoc computational creativity tool for making DS science pipelines' design friendly for non-data scientists.

\section{Towards a Human in the loop creative platform for designing data science pipelines}\label{sec:platform}

\begin{figure*}[t]
   \centering
\includegraphics[width=0.9\linewidth]{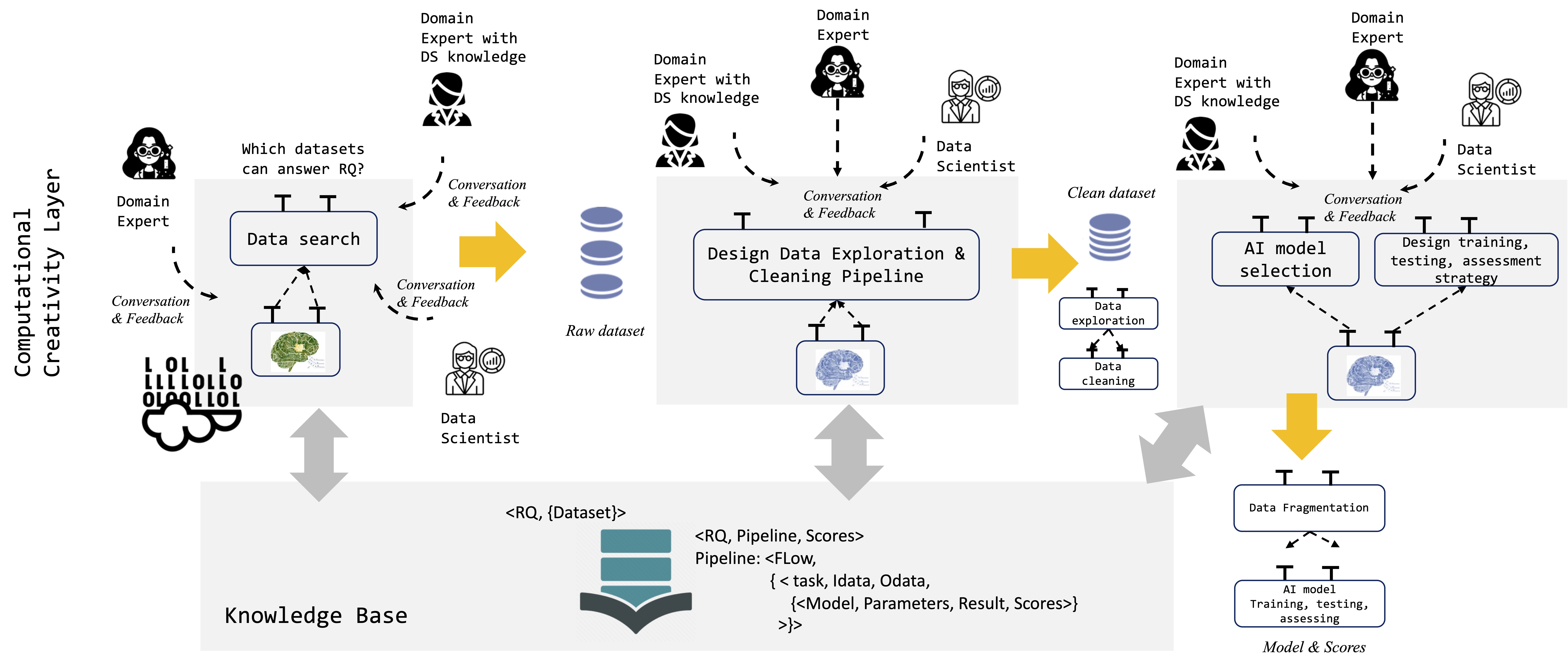}
 \caption{Matilda platform creation pipeline.}
   \label{fig:architecture}
 \end{figure*}

Figure \ref{fig:architecture} shows the general architecture of the MATILDA  platform that assists people with different expertise to follow a creative process for designing DS pipelines given datasets and target research questions. The platform relies on a step-by-step conversational approach based on our previous work \cite{bethaz2021ds4all} and provides interaction entry points to allow humans feedback, validate and guide the creative process. For each phase of a DS pipeline (data exploration and preparation, fragmentation, training, testing and assessing), the platform suggests possible scenarios that are adopted or not.


Our DS creativity platform allows us to study how overall creativity is affected if computer systems take over different roles within the design of data science pipelines. The platform provides a collaborative environment that integrates an artificial actor within 
in the creative production process  of
DS pipelines by data scientists.


\section{Conclusions and Future Work}\label{sec:conclusion}
The research and development market associated with data science is fuelling the economies of countries in the world. Almost all sectors in the global economies see data science as a promising alternative to develop original solutions to critical societal problems and promote data-driven decision-making processes that can create know-how and value. Yet, despite the availability of datasets and technology for transforming any phenomenon produced in reality into digital data and the variety of algorithms (Mathematical and artificial intelligence models), the design of data science solutions remains artisanal. The impact on person-hours and economic investment is not anecdotic. The time has come to propose methodologies that can formalise the design of data science solutions and model the “know-how” developed by data scientists during the creation process. Besides, data science addresses trans-disciplinary challenges. It is critical to bridge the gap between technical vocabulary, tasks, and the vocabulary of other disciplines and users with different expertise. This strategy will ensure the usability and acceptability of solutions (i.e. pipelines). In summary, data science must become inclusive and accessible to all. Our work addresses this challenge by aiming to adopt computational creativity methods to model the data science design process(es) that combine human and nonhuman creativity.

The platform MATILDA proposed in this paper is based on the original methodologies that we propose. It contributes to creating data science pipelines according to the expectations of knowledge discovery. It is interesting for answering target research questions, the input data's characteristics and the data scientists' models. Creativity-based methodologies applied to data science will make it accessible and inclusive to address increasingly complex problems humanity faces.

\bibliographystyle{aaai}
\bibliography{biblio}

\end{document}